\documentclass[aps,pra,10pt,twocolumn,showkeys,nofootinbib,showpacs,superscriptaddress]{revtex4-1}

\usepackage{graphicx}
\usepackage{url}
\usepackage{amsmath}
\usepackage{amssymb}
\usepackage{bm}
\usepackage{dsfont}
\usepackage{graphicx}
\usepackage{subfigure}
\usepackage{dcolumn}
\usepackage{braket}
\usepackage{bm}
\usepackage{bbm}
\usepackage{color}

\newcommand{\imag}{\mathrm{i}}
\newcommand{\ex}{\mathrm{e}}
\newcommand{\pder}[2][]{\frac{\partial#1}{\partial#2}}

\newcommand{\idm}{\mathbbm{1}}
\newcommand{\pauliX}{\hat{\sigma}_{\mathrm{x}}}
\newcommand{\pauliY}{\hat{\sigma}_{\mathrm{y}}}
\newcommand{\pauliZ}{\hat{\sigma}_{\mathrm{z}}}
\newcommand{\vecA}[1]{\hat{\mathbf{A}}_{#1}^{\vphantom{+}}}
\newcommand{\vecAd}[1]{\hat{\mathbf{A}}_{#1}^{+}}
\newcommand{\vecC}[1]{\hat{\mathbf{C}}_{#1}^{\vphantom{+}}}
\newcommand{\vecCd}[1]{\hat{\mathbf{C}}_{#1}^{+}}
\newcommand{\vecMC}[1]{\hat{\bm{\mathcal{C}}}_{#1}^{\vphantom{+}}}
\newcommand{\vecMCd}[1]{\hat{\bm{\mathcal{C}}}_{#1}^{+}}

\begin{document}

\title{Spin-Orbit Coupling in Periodically Driven Optical Lattices}

\author{J. Struck}
\email[E-mail address: ]{jstruck@mit.edu}
\altaffiliation[\\ Present address: ]
{MIT-Harvard Center for Ultracold Atoms, Research Laboratory of Electronics, and Department of Physics,
Massachusetts Institute of Technology, Cambridge, Massachusetts 02139, USA}
\affiliation{Institut f\"ur Laserphysik, Universit\"at Hamburg, Luruper Chaussee 149, D-22761 Hamburg, Germany}

\author{J. Simonet}
\author{K. Sengstock}
\affiliation{Institut f\"ur Laserphysik, Universit\"at Hamburg, Luruper Chaussee 149, D-22761 Hamburg, Germany}
\affiliation{ZOQ, Universit\"at Hamburg, Luruper Chaussee 149, D-22761 Hamburg, Germany}

\begin{abstract}
We propose a method for the emulation of artificial spin-orbit coupling in a system of ultracold, neutral atoms trapped in a tight-binding lattice. This scheme does not involve near-resonant laser fields, avoiding the heating processes connected to the spontaneous emission of photons. In our case, the necessary spin-dependent tunnel matrix elements are generated by a rapid, spin-dependent, periodic force, which can be described in the framework of an effective, time-averaged Hamiltonian. An additional radio-frequency coupling between the spin states leads to a mixing of the spin bands.
\end{abstract}


\maketitle

The phenomenon of spin-orbit coupling (SOC) generally describes an interplay between the spin state of a particle and its motional degrees of freedom. This effect naturally arises in the framework of relativistic quantum mechanics described by the Dirac equation. A spinful particle moving through an electric field experiences a magnetic field in the co-moving reference frame. The resulting interaction between the spin and the magnetic field depends on the amplitude of the field and thus on the velocity of the particle, leading to the coupling of motion and spin. In solid state materials, SOC can result in exotic phases and phenomena, such as topological insulators \cite{Hsieh:2008ie,Hasan:2010ku} or the spin Hall effect \cite{Dyakonov:1971ml,D'Yakonov:1971jm,Hirsch:1999hn,Kato:2004ft,Wunderlich:2005bk}.

The experimental realization of synthetic spin-orbit interactions with equal Rashba \cite{Bychkov:1984ek} and Dresselhaus \cite{Dresselhaus:1955en} contributions for bosonic \cite{Lin:2011hn,Zhang:2012fd} and fermionic quantum gases \cite{Cheuk:2012id,Wang:2012gv} has raised considerable interest over the last years (\cite{Galitski:2013dh} and references therein). For ultracold atoms -- interacting via s-wave scattering -- SOC can lead to effective interactions with higher partial wave contributions \cite{Williams:2012gs,Williams:2013,Fu:2013bq}. In a single-component Fermi gas this could lead to stable p-wave interactions and topological superfluids \cite{Massignan:2010hm,Jiang:2011cw,Seo:2012gn}. All of the experimentally implemented schemes rely on near-resonant Raman-laser coupling schemes and thus suffer from spontaneous emission, leading to excitations and particle loss. Currently a lot of effort is directed towards novel methods for the creation of artificial SOC avoiding this issue, e.g., by using magnetic field pulses \cite{Anderson:2013,Xu:2013} or far-off-resonant light \cite{Kennedy:2013fp}.

As thoroughly investigated over the last years a rapid, periodic lattice drive coherently manipulates the tunneling processes in an optical lattice. This allows for, e.g., the coherent destruction of tunneling and sign inversion of the tunnel elements \cite{Eckardt:2005bq,Lignier:2007du,Kierig:2008kb,Zenesini:2009ie}, photon assisted tunneling \cite{Eckardt:2007it,Sias:2008jf}, and the creation of artificial gauge fields \cite{Struck:2011,Struck:2012gc,Hauke:2012dh,Struck:2013ar,Jotzu:2014,Zhang:2014,Goldman:2014,Baur:2014,Kosior:2014}.
\begin{figure}
\centering
\includegraphics[width=8.6cm]{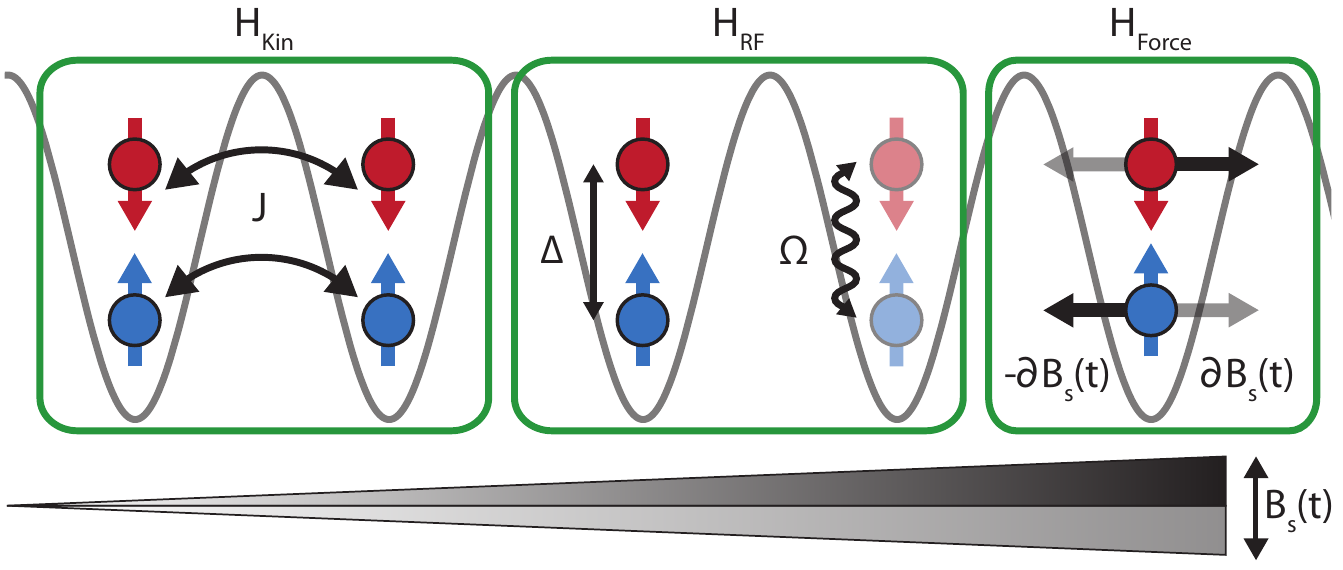}
\caption{Schematic illustration of the relevant processes, which are captured in the two-component single-particle Hamiltonian. $H_{\mathrm{Kin}}$ represents the spin-independent nearest-neighbour tunneling processes, $H_{\mathrm{RF}}$ describes the radio frequency coupling of the two spin states and $H_{\mathrm{Force}}$ corresponds to the spin-dependent driving of the atoms in the lattice. The drive originates from an oscillating magnetic field gradient.}
\label{fig:sos SpinOrbitBasicTerm}
\end{figure}
In this manuscript, we propose to engineer a time-periodic and spin-dependent drive in order to realize spin-dependent tunnel matrix elements and SOC. We consider ultracold atoms with (at least) two spin states confined in a 1D optical lattice. A time-dependent magnetic field $(B_{s}(t)=B_{s}(t+T))$ periodically drives the atoms at the frequency $\omega=2\pi/T$. For simplicity we assume that the two relevant spin states are characterized by magnetic moments with same magnitude but opposite signs so that they are submitted to opposite forces. However all results presented here can be generalized to spin combinations experiencing a different drive.

A time-periodic driving can lead to complex valued tunnel matrix elements if the driving breaks specific symmetries, resulting in a gauge dependent shift of the dispersion relation for a 1D lattice \cite{Struck:2012gc}. In our case, the dispersion relations of the two spin states are shifted in opposite direction due to the inverted drive for both states. An additional radio-frequency coupling between the spin states leads to a mixing of the spin dispersion relations and a spin-orbit gap in the band structure.

In this scheme, the strength of the SOC can be continuously tuned, simply by adjusting the driving amplitude (see Ref. \cite{Zhang:2013es} for a Raman coupling scheme with tunable SOC strength). Importantly as no near-resonant laser fields are involved, the quantum gas is not prone to heating processes induced by the spontaneous emission of photons.

The Hamiltonian can be separated into three terms $H = \,H_{\mathrm{Kin}} + H_{\mathrm{RF}} + H_{\mathrm{Force}}$ (Fig. \ref{fig:sos SpinOrbitBasicTerm}). Here we neglected interactions between the particles. However, we would like to emphasize that spin preserving, on-site Hubbard type interactions are not renormalized by the drive in the effective Hamiltonian.

First, $H_{\mathrm{Kin}}$ describes the next-neighbor tunneling processes, which are spin-inde\-pendent and do not couple the spin states:
\begin{equation}
H_{\mathrm{Kin}} = -J \sum_{s} \left( \vecAd{s} \vecA{s-1} +  \vecAd{s-1} \vecA{s} \right),
\end{equation}
where $J$ is the tunnel matrix element between neighbouring lattice sites and $\vecA{s}=({\hat{a}_{s}},{\hat{b}_{s}})^{\mathrm{T}}$, $\vecAd{s}=({\hat{a}_{s}}^{+},{\hat{b}_{s}}^{+})$ are vectors with the annihilation operators $\hat{a}_{s}$ and $\hat{b}_{s}$ on lattice site $s$ for the two spin components.
For the rest of this manuscript we assume that these operators obey the bosonic commutation relations. Nevertheless, the single-particle results derived here are valid for fermionic particles as well.

Second, the Rabi coupling of both spin states in the rotating-wave approximation is described by
\begin{equation}
H_{\mathrm{RF}}  = \frac{\hbar \Delta}{2} \sum_{s} \vecAd{s} \pauliZ  \vecA{s} - \frac{\hbar \Omega}{2} \sum_{s} \vecAd{s} \pauliX \vecA{s},
\end{equation}
where $\Delta= \omega_{\mathrm{A}}-\omega_{\mathrm{RF}}$ is the detuning of the radio-wave ($\omega_{\mathrm{RF}}$) with respect to the atomic resonance ($\omega_{\mathrm{A}}$), $\Omega$ is the Rabi frequency and $\hat{\sigma}_{\mathrm{x},\mathrm{y},\mathrm{z}}$ denote the Pauli matrices.

The third term describes the spin-dependent driving of the atoms by an oscillating magnetic field gradient:
\begin{equation}
H_{\mathrm{Force}} =  \sum_{s} v_{s}(t) \vecAd{s} \pauliZ  \vecA{s}. \label{eq:driving term Hamiltonian}
\end{equation}
Here $v_{s}(t) = \mathbf{\mu} B_{s}(t) $ is the energy shift due to the presence of the oscillating magnetic field. The resulting force is proportional to the gradient of the field and acts in opposite directions for the two different spin components.

The time-periodic Hamiltonian $H$ can be treated in the framework of the Floquet theory, where the energy structure is described by a time-independent, periodic quasi-energy spectrum \cite{Shirley:1965cy,Sambe:1973hi}. In the high driving frequency limit -- where the coupling elements between different Floquet bands can be neglected -- the quasi-energy spectrum can be approximated by an effective Hamiltonian \cite{Eckardt:2005bq, Hemmerich:2010fh}.
The effective Hamiltonian $H_{\mathrm{eff}}$ can be obtained from the Floquet Hamiltonian $H_{\mathcal{F}} = H - \imag \hbar \pder{t}$ by the relation
\begin{equation}
H_{\mathrm{eff}} =  \Braket{ \hat{U}_{Q}^{+}(t) H_{\mathcal{F}} \hat{U}_{Q}(t)}_{T}, \label{eq:effective Hamiltonian general expression}
\end{equation}
with the unitary operator $\hat{U}_{Q}(t) = \exp(-\imag \hat{Q}(t))$ and the notation $\braket {\cdots }_{T} = \frac{1}{T} \int_{0}^{T} \cdots \mathrm{d} t$ for the time-average. The hermitian, time-periodic operator $\hat{Q}(t)$ is chosen such that the driving term (Eq. \eqref{eq:driving term Hamiltonian}) is absent in the effective Hamiltonian \cite{Eckardt:2005bq}:
\begin{equation}
\hat{Q} = \frac{1}{\hbar}  \sum_{s} W_{s}(t) ~ \vecAd{s} \pauliZ \vecA{s}.
\end{equation}
Here, $W_{s}(t)$ is defined by,
\begin{equation}
     W_{s}(t) = \int_{t_{0}}^{t} \mathrm{d} t' v_{s}(t') - \Braket{\int_{t_{0}}^{t} \mathrm{d} t' v_{s}(t')}_{T}, \label{eq:Ws}
\end{equation}
for times $t \geq t_{0}$, where $t_{0}$ denotes the time when the amplitude of the time-dependent magnetic field has been fully ramped up. Eq. \eqref{eq:Ws} is independent of $t_{0}$ for $t \geq t_{0}$.
The effective Hamiltonian (Eq. \eqref{eq:effective Hamiltonian general expression}) then reads
\begin{align}
H_{\mathrm{eff}} =&  -J \sum_{s} |f_{s}| \left( \vecAd{s} \ex^{\imag \theta_{s} \pauliZ} \vecA{s-1} +  \vecAd{s-1} \ex^{-\imag \theta_{s} \pauliZ} \vecA{s} \right) \nonumber
\\
&- \frac{\hbar \Omega}{2} \sum_{s} |g_{s}| \vecAd{s} \left( \cos(\chi_{s}) \pauliX - \sin(\chi_{s}) \pauliY \right) \vecA{s} \nonumber
\\
&+ \frac{\hbar \Delta}{2} \sum_{s} \vecAd{s} \pauliZ  \vecA{s}, \label{eq:effective Hamiltonian bare basis}
\end{align}
where the complex variables $f_{s}$ and $g_{s}$ have been decomposed into magnitude and phase ($f_{s}{=}|f_{s}|\exp(\imag \theta_{s})$, $g_{s} {=} |g_{s}|\exp(\imag \chi_{s})$).
Here, the function
\begin{equation}
f_{s} \equiv   \Braket{\exp \left(\imag \left[W_{s}(t) - W_{s-1}(t) \right]/\hbar \right)}_{T}, \label{eq:function f}
\end{equation}
describes the spin-dependent renormalization of the tunneling matrix elements, whereas
\begin{equation}
g_{s} \equiv  \Braket{\exp \left(\imag \, 2 W_{s}(t)/\hbar \right)}_{T}, \label{eq:function g}
\end{equation}
describes the renormalization of the Rabi frequency. In general, both functions depend on the lattice site index.

The spatial rotation of the Pauli matrices in the Rabi coupling term of Eq. \eqref{eq:effective Hamiltonian bare basis} can be canceled by the transformation into dressed spin states, which are defined by the operators $\vecC{s} = \hat{T}_{s}^{+} \vecA{s}$ and $\vecCd{s} =\vecAd{s} \hat{T}_{s}$. The unitary operator $\hat{T}_{s}=\exp(\imag \chi_{s} \hat{\sigma}_{z}/2)$ represents a local rotation in spin space around the z-axis.
In the new basis the effective Hamiltonian is given by
\begin{align}
H_{\mathrm{eff}} =&  -J \sum_{s} |f_{s}| \left( \vecCd{s} \ex^{\imag \alpha_{s} \pauliZ} \vecC{s-1} +  \vecCd{s-1} \ex^{-\imag \alpha_{s} \pauliZ} \vecC{s} \right) \nonumber
\\
&- \frac{\hbar \Omega}{2} \sum_{s} |g_{s}| \vecCd{s} \pauliX \vecC{s} + \frac{\hbar \Delta}{2} \sum_{s} \vecCd{s} \pauliZ  \vecC{s},
\label{eq:effective Hamiltonian dressed basis}
\end{align}
where we have introduced the SOC parameter $\alpha_{s}=\theta_{s} + (\chi_{s-1} - \chi_{s})/2$, which can be understood as a spin-dependent Peierls phase. The Hamiltonian \eqref{eq:effective Hamiltonian dressed basis} describes a tight-binding lattice with one-dimensional SOC.

Generally, the site dependence of the parameters $|f_{s}|$, $\alpha_{s}$ and $|g_{s}|$ (see Eq. \eqref{eq:effective Hamiltonian dressed basis}) breaks the translational symmetry of the lattice potential and prevents a meaningful description in terms of Bloch states. However, for certain driving functions $v_{s}(t)$ these parameters are in good approximation site-independent.

\begin{figure}
\centering
\includegraphics[width=8.6cm]{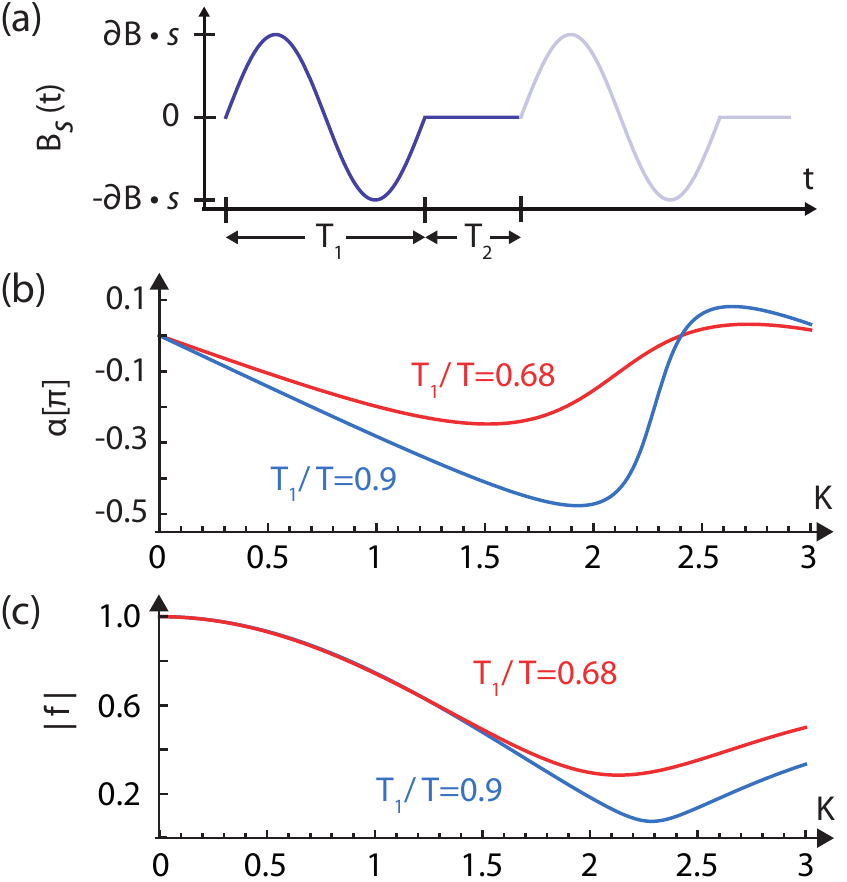}
\caption{Spin-dependent driving of the atoms with sinusoidal pulses. (a) The amplitude of the magnetic field is modulated around zero with trains of sinusoidal pulses. (b) The resulting SOC strength and (c) the renormalization  of the tunneling rate as functions of the forcing parameter $K$. In (b) and (c) the colored lines correspond to different ratios $T_{1}/T$ of the pulse to hold time.}
\label{fig:sine_pulse_alpha}
\end{figure}

This is for instance the case for a sinusoidally pulsed magnetic field given by (see Fig. \ref{fig:sine_pulse_alpha}(a)):
\begin{equation}
B_{s}(t) = \partial B \cdot s \, \begin{cases}
 \sin{(\omega_1 t)} & \text{for} ~ 0\phantom{_{1}}<t \text{ mod } T<T_1 ~ (\star)
    \\
  0 & \text{for} ~ T_1<t \text{ mod } T<T_{\phantom{1}} ~ (\star \star), \label{eq:time dependent magnetic field}
\end{cases}
\end{equation}
where $\partial B$ is the magnetic field difference between neighbouring lattice sites and $\omega_1 = 2\pi / T_{1}$. This function has two important features: First, it breaks time-reversal symmetry, leading to a continuously tunable SOC parameter. Second, it vanishes during a finite time interval $T_2$ ($T=T_1+T_2$), which is essential for a non-vanishing time-average of the Rabi coupling in the effective Hamiltonian. Insertion of Eq. \eqref{eq:time dependent magnetic field} into Eq. \eqref{eq:Ws} results in
\begin{equation}
W_{s}(t) = \hbar  \, K \, s \begin{cases}
 \cos{(\omega_1 t)} - T_{2}/T & ~ (\star)
    \\
  T_{1}/T & ~ (\star \star),
\end{cases} \label{eq:sinusoidal pulse Ws}
\end{equation}
with $K {=} \mu \, \partial B / \hbar \omega_1$ as the dimensionless forcing parameter.
The renormalization function of the tunnel matrix elements (Eq. \eqref{eq:function f})
\begin{equation}
f_{s} = \ex^{ - \imag K T_{2}/T} J_{0}^{\mathrm{B}}(K) \, T_{1}/T + \ex^{\imag K T_{1}/T}  T_{2}/T \equiv f(K),
\end{equation}
is completely site-independent, with $J_{0}^{\mathrm{B}}(K)$ as the zeroth order Bessel function of the first kind. In contrast the renormalization function of the Rabi frequency $g_{s}=f(2Ks)$ (see Eq. \eqref{eq:function g}) shows an explicit site dependence. However, in the limit $2 K s \rightarrow \infty $ this site dependence only affects the complex phase and we obtain:
\begin{equation}
g_{s} = \ex^{\imag 2 K s T_{1}/T} \, T_{2}/T. \label{eq:approximated function g}
\end{equation}
If the forcing parameter $K$ is on the order of one and the zero-crossing of the magnetic field is far away (typically a few thousands of lattice sites) from the center of the atomic cloud, then Eq. \eqref{eq:approximated function g} is a good approximation for the exact result of Eq. \eqref{eq:function g}

In this limit the magnitude $|g_{s}| \approx T_{2}/T \equiv |g|$ is site-independent.
Note that for $T_{2}=0$ the effective Rabi coupling between the spin states vanishes. The complex phase of $g_{s}$ is given by $\chi_{s} \approx 2 K s T_{1}/T$ and thus we obtain a site-independent SOC parameter $\alpha=\theta - K T_{1}/T$.
This parameter can be continuously tuned via the forcing parameter $K$ and the ratio $T_{1}/T$ as shown in Fig. 2(b). For larger $T_{1}/T$, the maximum value of $\alpha$ increases in contrast to the effective Rabi frequency $\Omega |g|$. The effective tunneling strength $J|f|$, reduced due to the periodic modulation, presents a minimum value which decreases to zero as the ratio $T_{1}/T$ is increased (Fig. 2(c)). Please note, the forcing parameter cannot become arbitrarily small as the condition $2 K s \rightarrow \infty $ needs to be fulfilled.

To estimate the strength of the required magnetic field gradient we assume to work with a bosonic alkali-metal with a nuclear spin of $I=3/2$ ($^{7}\mathrm{Li}$, $^{23}\mathrm{Na}$ or $^{87}\mathrm{Rb}$) in the low field Zeeman regime. As the two spin states we use the hyperfine states $\ket{F=1,m_{F}=-1}$ and $\ket{F=2,m_{F}=-1}$ of the $n^{2}S_{1/2}$ ground state. The required magnetic field gradient for a forcing parameter $K$ would be $\partial B/d = K \hbar \omega_1/(d \mu_{\mathrm{B}} m_{F} g_{F}) \approx K \cdot 31 \, \mathrm{G}/\mathrm{cm},$ where the Land\'{e} factor is given by $g_{F=1,2} = \mp 1/2$ and we have assumed a frequency of $\omega_1 = 2\pi \cdot 1.1 \mathrm{kHz}$. The lattice spacing $d$ has been set to $0.5 \mathrm{\mu m}$.

Although this scheme is free of heating arising from spontaneous emission of photons, the periodic driving can create excitations in the system. The heating rate strongly depends on the specific driving frequency, lattice depth and strength of possible interactions between the particles \cite{Eckardt:2005bq,Lignier:2007du,Zenesini:2009ie,Arlinghaus:2010bz,Poletti:2011}. Therefore the driving frequency has to be chosen out of resonance with the direct and multi-photon transitions to higher bands of the lattice. In addition the effective Hamiltonian derived here is only a valid approximation if the energy corrections due to the coupling between different Floquet bands are negligible. From perturbation theory follows that the Floquet band coupling elements $|\braket{\hat{V}_{p,m}}|$ and the eigenvalues of $H_{\mathrm{eff}}$ have to be small compared to the energetic separation $\hbar \omega$ between the Floquet bands. The coupling between two Floquet bands with indices $m$ and $p$ arises from the perturbation operator $\hat{V}_{p,m} = \braket{  \mathrm{e}^{\imag (p-m) \, \omega t} \hat{U}_{Q}^{+}(t) H_{\mathcal{F}} \hat{U}_{Q}(t)}_{T}$ \cite{Hemmerich:2010fh}. Evaluating the perturbation operator for the sinusoidal pulse drive, we arrive at the following constraints:
\begin{equation}
   |\xi_{p,m}(1)| \cdot J, ~ |\xi_{p,m}(2s)| \cdot \hbar\Omega \ll \hbar  \omega, \label{eq:lower frequency bound}
\end{equation}
with $\xi_{p,m}(x)= \braket{ \mathrm{e}^ {\imag [ (p-m) \omega t +  W_{x}(t)/\hbar ]}}_{T}$ and $W_{x}(t)$ given by Eq. \eqref{eq:sinusoidal pulse Ws}. The function $|\xi_{p,m}(x)|$ is always smaller than one and converges for large $x$ towards a value which never exceeds $|g|$. Therefore, a reduced effective Rabi frequency $|g| \Omega$ can always be compensated by increasing the bare Rabi frequency without violating the constraints given in \eqref{eq:lower frequency bound}. From the condition for the eigenvalues of the effective Hamiltonian we obtain the additional constraints:
\begin{equation}
   |g| \cdot J, ~ |f| \cdot \hbar\Omega, ~ \hbar \Delta \ll \hbar  \omega. \label{eq:lower frequency bound eigenvalues}
\end{equation}
Note that, for an interacting system the driving frequency has to be also large compared to the energy scale of the interactions. For the driving frequency, these constraints result in a lower bound given by the energy scale of the tunneling strength and an upper bound defined by the band gap. Furthermore, the driving frequency hast be chosen out resonance with multi-photon excitations lying in between the lower and upper bound.

\begin{figure}
\centering
\includegraphics[width=8.6cm]{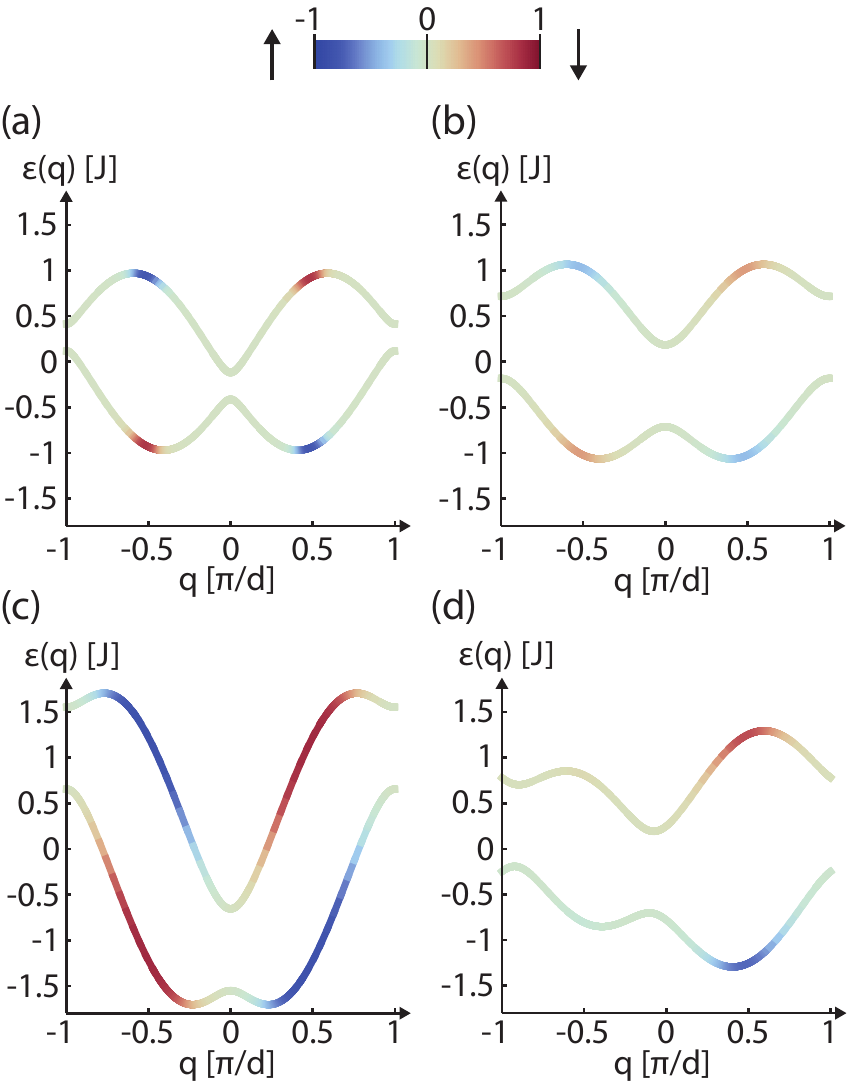}
\caption{Dispersion relations of the spin-orbit coupled system for sinusoidal pulses with $T_{1}/T=0.9$. The colorcode indicates the respective admixture of the bare spin states (Eq. \eqref{eq:bare spins reciprocal space}). The forcing parameter, Rabi frequency and detuning are (a) $K=1.5$, $\hbar \Omega = 3 J$ and $\hbar \Delta = 0$,  (b) $K=1.5$, $\hbar \Omega = 9 J$ and $\hbar \Delta = 0$, (c) $K=0.9$, $\hbar \Omega = 9 J$ and $\hbar \Delta = 0$ and (d) $K=1.5$, $\hbar \Omega = 9 J$ and $\hbar \Delta = 0.5 J$.}
\label{fig:SpinOrbitDispersion}
\end{figure}

The dispersion relation of the spin-orbit coupled lattice can be obtained by the transformation of the annihilation operator for a particle on site $s$ into the reciprocal space:
\begin{equation}
\vecC{s} = \frac{1}{\sqrt{M}} \sum_{q} \vecMC{q} \mathrm{e}^{\imag \, q \, d \, s}, \label{eq:spinor reciprocal space}
\end{equation}
where $M$ is the number of lattice sites and $\vecMC{q} = (\hat{\mathcal{C}}_{q}^{1},\hat{\mathcal{C}}_{q}^{2})^{\mathrm{T}}$ is the two-component annihilation operator for a particle in the Bloch state with quasimomentum $q$.
The transformation of the creation and annihilation operators in the Hamiltonian \eqref{eq:effective Hamiltonian dressed basis} into the reciprocal space (Eq. \eqref{eq:spinor reciprocal space}) leads to
\begin{equation}
H_{\mathrm{eff}} = \sum_{q} \vecMCd{q} \mathcal{H} (q) \, \vecMC{q} \, ,
\end{equation}
with the matrix
\begin{align}
\mathcal{H} (q) = &- 2 J |f| \left( \cos(\alpha) \cos(q d) \, \idm  + \sin(\alpha) \sin(q d) \, \pauliZ \right) \nonumber
\\
&+ \hbar \Delta \pauliZ / 2 - \hbar \Omega |g| \pauliX / 2 \,. \label{eq:hamiltonian reciprocal space}
\end{align}
The eigenvalues of the matrix $\mathcal{H} (q)$ represent the two lowest spinful bands of the spin-orbit coupled lattice (Fig.~\ref{fig:SpinOrbitDispersion}).
An increase of the Rabi frequency $\Omega$ leads to a larger splitting of these bands (Figs. \ref{fig:SpinOrbitDispersion}(a) and (b)). The forcing parameter $K$ directly influences the SOC parameter $\alpha$ and the bandwidth via the renormalization $|f|$ of the tunneling rate (Figs. \ref{fig:SpinOrbitDispersion}(b) and (c)). A finite detuning $\Delta$ lifts the reflection symmetry of the dispersion relation around quasimomentum zero (Fig. \ref{fig:SpinOrbitDispersion}(d)). Experimentally it is therefore necessary to reduce short-term magnetic field fluctuations, directly translating into a time-dependent detuning and heating, to an energy scale which is small compared to the effective tunneling strength. This issue can be partially resolved by using a light atomic species, where the energy scale connected to the tunneling is larger than for heavier elements. However, we would like to point out that this is a generic problem of all SOC schemes using Zeeman states with different magnetic moments.

The combination of time-of-flight absorption imaging and a Stern-Gerlach spin separation allows to probe the admixture of the bare spin components in the dispersion relation. The quasimomentum dependent admixture of the bare spin states is related to the transformed spin basis of Eq. \eqref{eq:hamiltonian reciprocal space} by
\begin{align}
\braket{\hat{\bm{\mathcal{A}}}_{q}^{+} \pauliZ  \hat{\bm{\mathcal{A}}}_{q}^{\vphantom{+}}} =  \hphantom{-} & \braket{{\hat{\mathcal{C}}_{q-K T_{1}/T}^{1^+}} {\hat{\mathcal{C}}_{q-K T_{1}/T}^{1}}}  \nonumber
\\
 - &\braket{{\hat{\mathcal{C}}_{q+K T_{1}/T}^{2^+}} {\hat{\mathcal{C}}_{q+K T_{1}/T}^{2}}}, \label{eq:bare spins reciprocal space}
\end{align}
where $\hat{\bm{\mathcal{A}}}_{q}^{\vphantom{+}}$ is the two-component annihilation operator for the bare spin states in the Bloch function with quasimomentum $q$. It can be obtained from $\vecA{s}$ by a Fourier expansion in analogy to Eq. \eqref{eq:spinor reciprocal space}.

In conclusion, we have proposed to use a periodically driven magnetic field gradient to create spin-dependent tunneling matrix elements for bosonic or fermionic species in an optical lattice. The Rabi coupling between different spin states leads to a gap in the dispersion relation and a mixing of the bare spin components. The proposed scheme avoids the problems associated with the spontaneous emission of photons. An additional advantage of our driving technique is the simple control over the SOC parameter $\alpha$, which can be easily tuned \emph{in-situ}.
Beyond that, the interplay between strong many-body interactions in an optical lattice -- not altered by the periodic drive -- and spin-dependent tunneling provides an ideal toolbox for the study of strongly correlated SOC systems \cite{Cole:2012,Zi:2012,Radic:2012,Xu:2014}.

We thank Andr\'{e} Eckardt, Christoph \"{O}lschl\"{a}ger and Malte Weinberg for stimulating and helpful discussions.
The work presented in this manuscript has been funded by the Deutsche Forschungsgemeinschaft (SFB 925) and by the EU under grant agreement No. 323714 (EQuaM).


\begin{thebibliography}{48}%
\makeatletter
\providecommand \@ifxundefined [1]{%
 \@ifx{#1\undefined}
}%
\providecommand \@ifnum [1]{%
 \ifnum #1\expandafter \@firstoftwo
 \else \expandafter \@secondoftwo
 \fi
}%
\providecommand \@ifx [1]{%
 \ifx #1\expandafter \@firstoftwo
 \else \expandafter \@secondoftwo
 \fi
}%
\providecommand \natexlab [1]{#1}%
\providecommand \enquote  [1]{``#1''}%
\providecommand \bibnamefont  [1]{#1}%
\providecommand \bibfnamefont [1]{#1}%
\providecommand \citenamefont [1]{#1}%
\providecommand \href@noop [0]{\@secondoftwo}%
\providecommand \href [0]{\begingroup \@sanitize@url \@href}%
\providecommand \@href[1]{\@@startlink{#1}\@@href}%
\providecommand \@@href[1]{\endgroup#1\@@endlink}%
\providecommand \@sanitize@url [0]{\catcode `\\12\catcode `\$12\catcode
  `\&12\catcode `\#12\catcode `\^12\catcode `\_12\catcode `\%12\relax}%
\providecommand \@@startlink[1]{}%
\providecommand \@@endlink[0]{}%
\providecommand \url  [0]{\begingroup\@sanitize@url \@url }%
\providecommand \@url [1]{\endgroup\@href {#1}{\urlprefix }}%
\providecommand \urlprefix  [0]{URL }%
\providecommand \Eprint [0]{\href }%
\providecommand \doibase [0]{http://dx.doi.org/}%
\providecommand \selectlanguage [0]{\@gobble}%
\providecommand \bibinfo  [0]{\@secondoftwo}%
\providecommand \bibfield  [0]{\@secondoftwo}%
\providecommand \translation [1]{[#1]}%
\providecommand \BibitemOpen [0]{}%
\providecommand \bibitemStop [0]{}%
\providecommand \bibitemNoStop [0]{.\EOS\space}%
\providecommand \EOS [0]{\spacefactor3000\relax}%
\providecommand \BibitemShut  [1]{\csname bibitem#1\endcsname}%
\let\auto@bib@innerbib\@empty
\bibitem [{\citenamefont {Hsieh}\ \emph {et~al.}(2008)\citenamefont {Hsieh},
  \citenamefont {Qian}, \citenamefont {Wray}, \citenamefont {Xia},
  \citenamefont {Hor}, \citenamefont {Cava},\ and\ \citenamefont
  {Hasan}}]{Hsieh:2008ie}%
  \BibitemOpen
  \bibfield  {author} {\bibinfo {author} {\bibfnamefont {D.}~\bibnamefont
  {Hsieh}}, \bibinfo {author} {\bibfnamefont {D.}~\bibnamefont {Qian}},
  \bibinfo {author} {\bibfnamefont {L.}~\bibnamefont {Wray}}, \bibinfo {author}
  {\bibfnamefont {Y.}~\bibnamefont {Xia}}, \bibinfo {author} {\bibfnamefont
  {Y.~S.}\ \bibnamefont {Hor}}, \bibinfo {author} {\bibfnamefont {R.~J.}\
  \bibnamefont {Cava}}, \ and\ \bibinfo {author} {\bibfnamefont {M.~Z.}\
  \bibnamefont {Hasan}},\ }\href {\doibase 10.1038/nature06843} {\bibfield
  {journal} {\bibinfo  {journal} {Nature}\ }\textbf {\bibinfo {volume} {452}},\
  \bibinfo {pages} {970} (\bibinfo {year} {2008})}\BibitemShut {NoStop}%
\bibitem [{\citenamefont {Hasan}\ and\ \citenamefont
  {Kane}(2010)}]{Hasan:2010ku}%
  \BibitemOpen
  \bibfield  {author} {\bibinfo {author} {\bibfnamefont {M.~Z.}\ \bibnamefont
  {Hasan}}\ and\ \bibinfo {author} {\bibfnamefont {C.~L.}\ \bibnamefont
  {Kane}},\ }\href {\doibase 10.1103/RevModPhys.82.3045} {\bibfield  {journal}
  {\bibinfo  {journal} {Rev. Mod. Phys.}\ }\textbf {\bibinfo {volume} {82}},\
  \bibinfo {pages} {3045} (\bibinfo {year} {2010})}\BibitemShut {NoStop}%
\bibitem [{\citenamefont {D'Yakonov}\ and\ \citenamefont
  {Perel'}(1971{\natexlab{a}})}]{Dyakonov:1971ml}%
  \BibitemOpen
  \bibfield  {author} {\bibinfo {author} {\bibfnamefont {M.~I.}\ \bibnamefont
  {D'Yakonov}}\ and\ \bibinfo {author} {\bibfnamefont {I.~V.}\ \bibnamefont
  {Perel'}},\ }\href@noop {} {\bibfield  {journal} {\bibinfo  {journal} {JETP
  Lett.}\ }\textbf {\bibinfo {volume} {13}},\ \bibinfo {pages} {467} (\bibinfo
  {year} {1971}{\natexlab{a}})}\BibitemShut {NoStop}%
\bibitem [{\citenamefont {D'Yakonov}\ and\ \citenamefont
  {Perel'}(1971{\natexlab{b}})}]{D'Yakonov:1971jm}%
  \BibitemOpen
  \bibfield  {author} {\bibinfo {author} {\bibfnamefont {M.~I.}\ \bibnamefont
  {D'Yakonov}}\ and\ \bibinfo {author} {\bibfnamefont {I.~V.}\ \bibnamefont
  {Perel'}},\ }\href {\doibase 10.1016/0375-9601(71)90196-4} {\bibfield
  {journal} {\bibinfo  {journal} {Phys. Lett. A}\ }\textbf {\bibinfo {volume}
  {35}},\ \bibinfo {pages} {459} (\bibinfo {year}
  {1971}{\natexlab{b}})}\BibitemShut {NoStop}%
\bibitem [{\citenamefont {Hirsch}(1999)}]{Hirsch:1999hn}%
  \BibitemOpen
  \bibfield  {author} {\bibinfo {author} {\bibfnamefont {J.~E.}\ \bibnamefont
  {Hirsch}},\ }\href {\doibase 10.1103/PhysRevLett.83.1834} {\bibfield
  {journal} {\bibinfo  {journal} {Phys. Rev. Lett.}\ }\textbf {\bibinfo
  {volume} {83}},\ \bibinfo {pages} {1834} (\bibinfo {year}
  {1999})}\BibitemShut {NoStop}%
\bibitem [{\citenamefont {Kato}\ \emph {et~al.}(2004)\citenamefont {Kato},
  \citenamefont {Myers}, \citenamefont {Gossard},\ and\ \citenamefont
  {Awschalom}}]{Kato:2004ft}%
  \BibitemOpen
  \bibfield  {author} {\bibinfo {author} {\bibfnamefont {Y.~K.}\ \bibnamefont
  {Kato}}, \bibinfo {author} {\bibfnamefont {R.~C.}\ \bibnamefont {Myers}},
  \bibinfo {author} {\bibfnamefont {A.~C.}\ \bibnamefont {Gossard}}, \ and\
  \bibinfo {author} {\bibfnamefont {D.~D.}\ \bibnamefont {Awschalom}},\ }\href
  {\doibase 10.1126/science.1105514} {\bibfield  {journal} {\bibinfo  {journal}
  {Science}\ }\textbf {\bibinfo {volume} {306}},\ \bibinfo {pages} {1910}
  (\bibinfo {year} {2004})}\BibitemShut {NoStop}%
\bibitem [{\citenamefont {Wunderlich}\ \emph {et~al.}(2005)\citenamefont
  {Wunderlich}, \citenamefont {Kaestner}, \citenamefont {Sinova},\ and\
  \citenamefont {Jungwirth}}]{Wunderlich:2005bk}%
  \BibitemOpen
  \bibfield  {author} {\bibinfo {author} {\bibfnamefont {J.}~\bibnamefont
  {Wunderlich}}, \bibinfo {author} {\bibfnamefont {B.}~\bibnamefont
  {Kaestner}}, \bibinfo {author} {\bibfnamefont {J.}~\bibnamefont {Sinova}}, \
  and\ \bibinfo {author} {\bibfnamefont {T.}~\bibnamefont {Jungwirth}},\ }\href
  {\doibase 10.1103/PhysRevLett.94.047204} {\bibfield  {journal} {\bibinfo
  {journal} {Phys. Rev. Lett.}\ }\textbf {\bibinfo {volume} {94}},\ \bibinfo
  {pages} {047204} (\bibinfo {year} {2005})}\BibitemShut {NoStop}%
\bibitem [{\citenamefont {Bychkov}\ and\ \citenamefont
  {Rashba}(1984)}]{Bychkov:1984ek}%
  \BibitemOpen
  \bibfield  {author} {\bibinfo {author} {\bibfnamefont {Y.~A.}\ \bibnamefont
  {Bychkov}}\ and\ \bibinfo {author} {\bibfnamefont {E.~I.}\ \bibnamefont
  {Rashba}},\ }\href {\doibase 10.1088/0022-3719/17/33/015} {\bibfield
  {journal} {\bibinfo  {journal} {J. Phys. C}\ }\textbf {\bibinfo {volume}
  {17}},\ \bibinfo {pages} {6039} (\bibinfo {year} {1984})}\BibitemShut
  {NoStop}%
\bibitem [{\citenamefont {Dresselhaus}(1955)}]{Dresselhaus:1955en}%
  \BibitemOpen
  \bibfield  {author} {\bibinfo {author} {\bibfnamefont {G.}~\bibnamefont
  {Dresselhaus}},\ }\href {\doibase 10.1103/PhysRev.100.580} {\bibfield
  {journal} {\bibinfo  {journal} {Phys. Rev.}\ }\textbf {\bibinfo {volume}
  {100}},\ \bibinfo {pages} {580} (\bibinfo {year} {1955})}\BibitemShut
  {NoStop}%
\bibitem [{\citenamefont {Lin}\ \emph {et~al.}(2011)\citenamefont {Lin},
  \citenamefont {Jim{\'e}nez-Garc{\'\i}a},\ and\ \citenamefont
  {Spielman}}]{Lin:2011hn}%
  \BibitemOpen
  \bibfield  {author} {\bibinfo {author} {\bibfnamefont {Y.~J.}\ \bibnamefont
  {Lin}}, \bibinfo {author} {\bibfnamefont {K.}~\bibnamefont
  {Jim{\'e}nez-Garc{\'\i}a}}, \ and\ \bibinfo {author} {\bibfnamefont {I.~B.}\
  \bibnamefont {Spielman}},\ }\href {\doibase 10.1038/nature09887} {\bibfield
  {journal} {\bibinfo  {journal} {Nature}\ }\textbf {\bibinfo {volume} {471}},\
  \bibinfo {pages} {83} (\bibinfo {year} {2011})}\BibitemShut {NoStop}%
\bibitem [{\citenamefont {Zhang}\ \emph {et~al.}(2012)\citenamefont {Zhang},
  \citenamefont {Ji}, \citenamefont {Chen}, \citenamefont {Zhang},
  \citenamefont {Du}, \citenamefont {Yan}, \citenamefont {Pan}, \citenamefont
  {Zhao}, \citenamefont {Deng}, \citenamefont {Zhai}, \citenamefont {Chen},\
  and\ \citenamefont {Pan}}]{Zhang:2012fd}%
  \BibitemOpen
  \bibfield  {author} {\bibinfo {author} {\bibfnamefont {J.-Y.}\ \bibnamefont
  {Zhang}}, \bibinfo {author} {\bibfnamefont {S.-C.}\ \bibnamefont {Ji}},
  \bibinfo {author} {\bibfnamefont {Z.}~\bibnamefont {Chen}}, \bibinfo {author}
  {\bibfnamefont {L.}~\bibnamefont {Zhang}}, \bibinfo {author} {\bibfnamefont
  {Z.-D.}\ \bibnamefont {Du}}, \bibinfo {author} {\bibfnamefont
  {B.}~\bibnamefont {Yan}}, \bibinfo {author} {\bibfnamefont {G.-S.}\
  \bibnamefont {Pan}}, \bibinfo {author} {\bibfnamefont {B.}~\bibnamefont
  {Zhao}}, \bibinfo {author} {\bibfnamefont {Y.-J.}\ \bibnamefont {Deng}},
  \bibinfo {author} {\bibfnamefont {H.}~\bibnamefont {Zhai}}, \bibinfo {author}
  {\bibfnamefont {S.}~\bibnamefont {Chen}}, \ and\ \bibinfo {author}
  {\bibfnamefont {J.-W.}\ \bibnamefont {Pan}},\ }\href {\doibase
  10.1103/PhysRevLett.109.115301} {\bibfield  {journal} {\bibinfo  {journal}
  {Phys. Rev. Lett.}\ }\textbf {\bibinfo {volume} {109}},\ \bibinfo {pages}
  {115301} (\bibinfo {year} {2012})}\BibitemShut {NoStop}%
\bibitem [{\citenamefont {Cheuk}\ \emph {et~al.}(2012)\citenamefont {Cheuk},
  \citenamefont {Sommer}, \citenamefont {Hadzibabic}, \citenamefont {Yefsah},
  \citenamefont {Bakr},\ and\ \citenamefont {Zwierlein}}]{Cheuk:2012id}%
  \BibitemOpen
  \bibfield  {author} {\bibinfo {author} {\bibfnamefont {L.~W.}\ \bibnamefont
  {Cheuk}}, \bibinfo {author} {\bibfnamefont {A.~T.}\ \bibnamefont {Sommer}},
  \bibinfo {author} {\bibfnamefont {Z.}~\bibnamefont {Hadzibabic}}, \bibinfo
  {author} {\bibfnamefont {T.}~\bibnamefont {Yefsah}}, \bibinfo {author}
  {\bibfnamefont {W.~S.}\ \bibnamefont {Bakr}}, \ and\ \bibinfo {author}
  {\bibfnamefont {M.~W.}\ \bibnamefont {Zwierlein}},\ }\href {\doibase
  10.1103/PhysRevLett.109.095302} {\bibfield  {journal} {\bibinfo  {journal}
  {Phys. Rev. Lett.}\ }\textbf {\bibinfo {volume} {109}},\ \bibinfo {pages}
  {095302} (\bibinfo {year} {2012})}\BibitemShut {NoStop}%
\bibitem [{\citenamefont {Wang}\ \emph {et~al.}(2012)\citenamefont {Wang},
  \citenamefont {Yu}, \citenamefont {Fu}, \citenamefont {Miao}, \citenamefont
  {Huang}, \citenamefont {Chai}, \citenamefont {Zhai},\ and\ \citenamefont
  {Zhang}}]{Wang:2012gv}%
  \BibitemOpen
  \bibfield  {author} {\bibinfo {author} {\bibfnamefont {P.}~\bibnamefont
  {Wang}}, \bibinfo {author} {\bibfnamefont {Z.-Q.}\ \bibnamefont {Yu}},
  \bibinfo {author} {\bibfnamefont {Z.}~\bibnamefont {Fu}}, \bibinfo {author}
  {\bibfnamefont {J.}~\bibnamefont {Miao}}, \bibinfo {author} {\bibfnamefont
  {L.}~\bibnamefont {Huang}}, \bibinfo {author} {\bibfnamefont
  {S.}~\bibnamefont {Chai}}, \bibinfo {author} {\bibfnamefont {H.}~\bibnamefont
  {Zhai}}, \ and\ \bibinfo {author} {\bibfnamefont {J.}~\bibnamefont {Zhang}},\
  }\href {\doibase 10.1103/PhysRevLett.109.095301} {\bibfield  {journal}
  {\bibinfo  {journal} {Phys. Rev. Lett.}\ }\textbf {\bibinfo {volume} {109}},\
  \bibinfo {pages} {095301} (\bibinfo {year} {2012})}\BibitemShut {NoStop}%
\bibitem [{\citenamefont {Galitski}\ and\ \citenamefont
  {Spielman}(2013)}]{Galitski:2013dh}%
  \BibitemOpen
  \bibfield  {author} {\bibinfo {author} {\bibfnamefont {V.}~\bibnamefont
  {Galitski}}\ and\ \bibinfo {author} {\bibfnamefont {I.~B.}\ \bibnamefont
  {Spielman}},\ }\href {\doibase 10.1038/nature11841} {\bibfield  {journal}
  {\bibinfo  {journal} {Nature}\ }\textbf {\bibinfo {volume} {494}},\ \bibinfo
  {pages} {49} (\bibinfo {year} {2013})}\BibitemShut {NoStop}%
\bibitem [{\citenamefont {Williams}\ \emph {et~al.}(2012)\citenamefont
  {Williams}, \citenamefont {Leblanc}, \citenamefont {Jimenez-Garcia},
  \citenamefont {Beeler}, \citenamefont {Perry}, \citenamefont {Phillips},\
  and\ \citenamefont {Spielman}}]{Williams:2012gs}%
  \BibitemOpen
  \bibfield  {author} {\bibinfo {author} {\bibfnamefont {R.~A.}\ \bibnamefont
  {Williams}}, \bibinfo {author} {\bibfnamefont {L.~J.}\ \bibnamefont
  {Leblanc}}, \bibinfo {author} {\bibfnamefont {K.}~\bibnamefont
  {Jimenez-Garcia}}, \bibinfo {author} {\bibfnamefont {M.~C.}\ \bibnamefont
  {Beeler}}, \bibinfo {author} {\bibfnamefont {A.~R.}\ \bibnamefont {Perry}},
  \bibinfo {author} {\bibfnamefont {W.~D.}\ \bibnamefont {Phillips}}, \ and\
  \bibinfo {author} {\bibfnamefont {I.~B.}\ \bibnamefont {Spielman}},\ }\href
  {\doibase 10.1126/science.1212652} {\bibfield  {journal} {\bibinfo  {journal}
  {Science}\ }\textbf {\bibinfo {volume} {335}},\ \bibinfo {pages} {314}
  (\bibinfo {year} {2012})}\BibitemShut {NoStop}%
\bibitem [{\citenamefont {Williams}\ \emph {et~al.}(2013)\citenamefont
  {Williams}, \citenamefont {Beeler}, \citenamefont {LeBlanc}, \citenamefont
  {Jim\'enez-Garc\'ia},\ and\ \citenamefont {Spielman}}]{Williams:2013}%
  \BibitemOpen
  \bibfield  {author} {\bibinfo {author} {\bibfnamefont {R.~A.}\ \bibnamefont
  {Williams}}, \bibinfo {author} {\bibfnamefont {M.~C.}\ \bibnamefont
  {Beeler}}, \bibinfo {author} {\bibfnamefont {L.~J.}\ \bibnamefont {LeBlanc}},
  \bibinfo {author} {\bibfnamefont {K.}~\bibnamefont {Jim\'enez-Garc\'ia}}, \
  and\ \bibinfo {author} {\bibfnamefont {I.~B.}\ \bibnamefont {Spielman}},\
  }\href {\doibase 10.1103/PhysRevLett.111.095301} {\bibfield  {journal}
  {\bibinfo  {journal} {Phys. Rev. Lett.}\ }\textbf {\bibinfo {volume} {111}},\
  \bibinfo {pages} {095301} (\bibinfo {year} {2013})}\BibitemShut {NoStop}%
\bibitem [{\citenamefont {Fu}\ \emph {et~al.}(2013)\citenamefont {Fu},
  \citenamefont {Huang}, \citenamefont {Meng}, \citenamefont {Wang},
  \citenamefont {Zhang}, \citenamefont {Zhang}, \citenamefont {Zhai},
  \citenamefont {Zhang},\ and\ \citenamefont {Zhang}}]{Fu:2013bq}%
  \BibitemOpen
  \bibfield  {author} {\bibinfo {author} {\bibfnamefont {Z.}~\bibnamefont
  {Fu}}, \bibinfo {author} {\bibfnamefont {L.}~\bibnamefont {Huang}}, \bibinfo
  {author} {\bibfnamefont {Z.}~\bibnamefont {Meng}}, \bibinfo {author}
  {\bibfnamefont {P.}~\bibnamefont {Wang}}, \bibinfo {author} {\bibfnamefont
  {L.}~\bibnamefont {Zhang}}, \bibinfo {author} {\bibfnamefont
  {S.}~\bibnamefont {Zhang}}, \bibinfo {author} {\bibfnamefont
  {H.}~\bibnamefont {Zhai}}, \bibinfo {author} {\bibfnamefont {P.}~\bibnamefont
  {Zhang}}, \ and\ \bibinfo {author} {\bibfnamefont {J.}~\bibnamefont
  {Zhang}},\ }\href {\doibase 10.1038/nphys2824} {\bibfield  {journal}
  {\bibinfo  {journal} {Nat. Phys.}\ }\textbf {\bibinfo {volume} {10}},\
  \bibinfo {pages} {110} (\bibinfo {year} {2013})}\BibitemShut {NoStop}%
\bibitem [{\citenamefont {Massignan}\ \emph {et~al.}(2010)\citenamefont
  {Massignan}, \citenamefont {Sanpera},\ and\ \citenamefont
  {Lewenstein}}]{Massignan:2010hm}%
  \BibitemOpen
  \bibfield  {author} {\bibinfo {author} {\bibfnamefont {P.}~\bibnamefont
  {Massignan}}, \bibinfo {author} {\bibfnamefont {A.}~\bibnamefont {Sanpera}},
  \ and\ \bibinfo {author} {\bibfnamefont {M.}~\bibnamefont {Lewenstein}},\
  }\href {\doibase 10.1103/PhysRevA.81.031607} {\bibfield  {journal} {\bibinfo
  {journal} {Phys. Rev. A}\ }\textbf {\bibinfo {volume} {81}},\ \bibinfo
  {pages} {031607} (\bibinfo {year} {2010})}\BibitemShut {NoStop}%
\bibitem [{\citenamefont {Jiang}\ \emph {et~al.}(2011)\citenamefont {Jiang},
  \citenamefont {Kitagawa}, \citenamefont {Alicea}, \citenamefont {Akhmerov},
  \citenamefont {Pekker}, \citenamefont {Refael}, \citenamefont {Cirac},
  \citenamefont {Demler}, \citenamefont {Lukin},\ and\ \citenamefont
  {Zoller}}]{Jiang:2011cw}%
  \BibitemOpen
  \bibfield  {author} {\bibinfo {author} {\bibfnamefont {L.}~\bibnamefont
  {Jiang}}, \bibinfo {author} {\bibfnamefont {T.}~\bibnamefont {Kitagawa}},
  \bibinfo {author} {\bibfnamefont {J.}~\bibnamefont {Alicea}}, \bibinfo
  {author} {\bibfnamefont {A.~R.}\ \bibnamefont {Akhmerov}}, \bibinfo {author}
  {\bibfnamefont {D.}~\bibnamefont {Pekker}}, \bibinfo {author} {\bibfnamefont
  {G.}~\bibnamefont {Refael}}, \bibinfo {author} {\bibfnamefont {J.~I.}\
  \bibnamefont {Cirac}}, \bibinfo {author} {\bibfnamefont {E.}~\bibnamefont
  {Demler}}, \bibinfo {author} {\bibfnamefont {M.~D.}\ \bibnamefont {Lukin}}, \
  and\ \bibinfo {author} {\bibfnamefont {P.}~\bibnamefont {Zoller}},\ }\href
  {\doibase 10.1103/PhysRevLett.106.220402} {\bibfield  {journal} {\bibinfo
  {journal} {Phys. Rev. Lett.}\ }\textbf {\bibinfo {volume} {106}},\ \bibinfo
  {pages} {220402} (\bibinfo {year} {2011})}\BibitemShut {NoStop}%
\bibitem [{\citenamefont {Seo}\ \emph {et~al.}(2012)\citenamefont {Seo},
  \citenamefont {Han},\ and\ \citenamefont {S{\'a}~de Melo}}]{Seo:2012gn}%
  \BibitemOpen
  \bibfield  {author} {\bibinfo {author} {\bibfnamefont {K.}~\bibnamefont
  {Seo}}, \bibinfo {author} {\bibfnamefont {L.}~\bibnamefont {Han}}, \ and\
  \bibinfo {author} {\bibfnamefont {C.~A.~R.}\ \bibnamefont {S{\'a}~de Melo}},\
  }\href {\doibase 10.1103/PhysRevLett.109.105303} {\bibfield  {journal}
  {\bibinfo  {journal} {Phys. Rev. Lett.}\ }\textbf {\bibinfo {volume} {109}},\
  \bibinfo {pages} {105303} (\bibinfo {year} {2012})}\BibitemShut {NoStop}%
\bibitem [{\citenamefont {Anderson}\ \emph {et~al.}(2013)\citenamefont
  {Anderson}, \citenamefont {Spielman},\ and\ \citenamefont
  {Juzeli{\=u}nas}}]{Anderson:2013}%
  \BibitemOpen
  \bibfield  {author} {\bibinfo {author} {\bibfnamefont {B.~M.}\ \bibnamefont
  {Anderson}}, \bibinfo {author} {\bibfnamefont {I.~B.}\ \bibnamefont
  {Spielman}}, \ and\ \bibinfo {author} {\bibfnamefont {G.}~\bibnamefont
  {Juzeli{\=u}nas}},\ }\href {\doibase 10.1103/PhysRevLett.111.125301}
  {\bibfield  {journal} {\bibinfo  {journal} {Phys. Rev. Lett.}\ }\textbf
  {\bibinfo {volume} {111}},\ \bibinfo {pages} {125301} (\bibinfo {year}
  {2013})}\BibitemShut {NoStop}%
\bibitem [{\citenamefont {Xu}\ \emph {et~al.}(2013)\citenamefont {Xu},
  \citenamefont {You},\ and\ \citenamefont {Ueda}}]{Xu:2013}%
  \BibitemOpen
  \bibfield  {author} {\bibinfo {author} {\bibfnamefont {Z.-F.}\ \bibnamefont
  {Xu}}, \bibinfo {author} {\bibfnamefont {L.}~\bibnamefont {You}}, \ and\
  \bibinfo {author} {\bibfnamefont {M.}~\bibnamefont {Ueda}},\ }\href {\doibase
  10.1103/PhysRevA.87.063634} {\bibfield  {journal} {\bibinfo  {journal} {Phys.
  Rev. A}\ }\textbf {\bibinfo {volume} {87}},\ \bibinfo {pages} {063634}
  (\bibinfo {year} {2013})}\BibitemShut {NoStop}%
\bibitem [{\citenamefont {Kennedy}\ \emph {et~al.}(2013)\citenamefont
  {Kennedy}, \citenamefont {Siviloglou}, \citenamefont {Miyake}, \citenamefont
  {Burton},\ and\ \citenamefont {Ketterle}}]{Kennedy:2013fp}%
  \BibitemOpen
  \bibfield  {author} {\bibinfo {author} {\bibfnamefont {C.~J.}\ \bibnamefont
  {Kennedy}}, \bibinfo {author} {\bibfnamefont {G.~A.}\ \bibnamefont
  {Siviloglou}}, \bibinfo {author} {\bibfnamefont {H.}~\bibnamefont {Miyake}},
  \bibinfo {author} {\bibfnamefont {W.~C.}\ \bibnamefont {Burton}}, \ and\
  \bibinfo {author} {\bibfnamefont {W.}~\bibnamefont {Ketterle}},\ }\href
  {\doibase 10.1103/PhysRevLett.111.225301} {\bibfield  {journal} {\bibinfo
  {journal} {Phys. Rev. Lett.}\ }\textbf {\bibinfo {volume} {111}},\ \bibinfo
  {pages} {225301} (\bibinfo {year} {2013})}\BibitemShut {NoStop}%
\bibitem [{\citenamefont {Eckardt}\ \emph {et~al.}(2005)\citenamefont
  {Eckardt}, \citenamefont {Weiss},\ and\ \citenamefont
  {Holthaus}}]{Eckardt:2005bq}%
  \BibitemOpen
  \bibfield  {author} {\bibinfo {author} {\bibfnamefont {A.}~\bibnamefont
  {Eckardt}}, \bibinfo {author} {\bibfnamefont {C.}~\bibnamefont {Weiss}}, \
  and\ \bibinfo {author} {\bibfnamefont {M.}~\bibnamefont {Holthaus}},\ }\href
  {\doibase 10.1103/PhysRevLett.95.260404} {\bibfield  {journal} {\bibinfo
  {journal} {Phys. Rev. Lett.}\ }\textbf {\bibinfo {volume} {95}},\ \bibinfo
  {pages} {260404} (\bibinfo {year} {2005})}\BibitemShut {NoStop}%
\bibitem [{\citenamefont {Lignier}\ \emph {et~al.}(2007)\citenamefont
  {Lignier}, \citenamefont {Sias}, \citenamefont {Ciampini}, \citenamefont
  {Singh}, \citenamefont {Zenesini}, \citenamefont {Morsch},\ and\
  \citenamefont {Arimondo}}]{Lignier:2007du}%
  \BibitemOpen
  \bibfield  {author} {\bibinfo {author} {\bibfnamefont {H.}~\bibnamefont
  {Lignier}}, \bibinfo {author} {\bibfnamefont {C.}~\bibnamefont {Sias}},
  \bibinfo {author} {\bibfnamefont {D.}~\bibnamefont {Ciampini}}, \bibinfo
  {author} {\bibfnamefont {Y.~P.}\ \bibnamefont {Singh}}, \bibinfo {author}
  {\bibfnamefont {A.}~\bibnamefont {Zenesini}}, \bibinfo {author}
  {\bibfnamefont {O.}~\bibnamefont {Morsch}}, \ and\ \bibinfo {author}
  {\bibfnamefont {E.}~\bibnamefont {Arimondo}},\ }\href {\doibase
  10.1103/PhysRevLett.99.220403} {\bibfield  {journal} {\bibinfo  {journal}
  {Phys. Rev. Lett.}\ }\textbf {\bibinfo {volume} {99}},\ \bibinfo {pages}
  {220403} (\bibinfo {year} {2007})}\BibitemShut {NoStop}%
\bibitem [{\citenamefont {Kierig}\ \emph {et~al.}(2008)\citenamefont {Kierig},
  \citenamefont {Schnorrberger}, \citenamefont {Schietinger}, \citenamefont
  {Tomkovic},\ and\ \citenamefont {Oberthaler}}]{Kierig:2008kb}%
  \BibitemOpen
  \bibfield  {author} {\bibinfo {author} {\bibfnamefont {E.}~\bibnamefont
  {Kierig}}, \bibinfo {author} {\bibfnamefont {U.}~\bibnamefont
  {Schnorrberger}}, \bibinfo {author} {\bibfnamefont {A.}~\bibnamefont
  {Schietinger}}, \bibinfo {author} {\bibfnamefont {J.}~\bibnamefont
  {Tomkovic}}, \ and\ \bibinfo {author} {\bibfnamefont {M.~K.}\ \bibnamefont
  {Oberthaler}},\ }\href {\doibase 10.1103/PhysRevLett.100.190405} {\bibfield
  {journal} {\bibinfo  {journal} {Phys. Rev. Lett.}\ }\textbf {\bibinfo
  {volume} {100}},\ \bibinfo {pages} {190405} (\bibinfo {year}
  {2008})}\BibitemShut {NoStop}%
\bibitem [{\citenamefont {Zenesini}\ \emph {et~al.}(2009)\citenamefont
  {Zenesini}, \citenamefont {Lignier}, \citenamefont {Ciampini}, \citenamefont
  {Morsch},\ and\ \citenamefont {Arimondo}}]{Zenesini:2009ie}%
  \BibitemOpen
  \bibfield  {author} {\bibinfo {author} {\bibfnamefont {A.}~\bibnamefont
  {Zenesini}}, \bibinfo {author} {\bibfnamefont {H.}~\bibnamefont {Lignier}},
  \bibinfo {author} {\bibfnamefont {D.}~\bibnamefont {Ciampini}}, \bibinfo
  {author} {\bibfnamefont {O.}~\bibnamefont {Morsch}}, \ and\ \bibinfo {author}
  {\bibfnamefont {E.}~\bibnamefont {Arimondo}},\ }\href {\doibase
  10.1103/PhysRevLett.102.100403} {\bibfield  {journal} {\bibinfo  {journal}
  {Phys. Rev. Lett.}\ }\textbf {\bibinfo {volume} {102}},\ \bibinfo {pages}
  {100403} (\bibinfo {year} {2009})}\BibitemShut {NoStop}%
\bibitem [{\citenamefont {Eckardt}\ and\ \citenamefont
  {Holthaus}(2007)}]{Eckardt:2007it}%
  \BibitemOpen
  \bibfield  {author} {\bibinfo {author} {\bibfnamefont {A.}~\bibnamefont
  {Eckardt}}\ and\ \bibinfo {author} {\bibfnamefont {M.}~\bibnamefont
  {Holthaus}},\ }\href {\doibase 10.1209/0295-5075/80/50004} {\bibfield
  {journal} {\bibinfo  {journal} {Europhys. Lett.}\ }\textbf {\bibinfo {volume}
  {80}},\ \bibinfo {pages} {50004} (\bibinfo {year} {2007})}\BibitemShut
  {NoStop}%
\bibitem [{\citenamefont {Sias}\ \emph {et~al.}(2008)\citenamefont {Sias},
  \citenamefont {Lignier}, \citenamefont {Singh}, \citenamefont {Zenesini},
  \citenamefont {Ciampini}, \citenamefont {Morsch},\ and\ \citenamefont
  {Arimondo}}]{Sias:2008jf}%
  \BibitemOpen
  \bibfield  {author} {\bibinfo {author} {\bibfnamefont {C.}~\bibnamefont
  {Sias}}, \bibinfo {author} {\bibfnamefont {H.}~\bibnamefont {Lignier}},
  \bibinfo {author} {\bibfnamefont {Y.~P.}\ \bibnamefont {Singh}}, \bibinfo
  {author} {\bibfnamefont {A.}~\bibnamefont {Zenesini}}, \bibinfo {author}
  {\bibfnamefont {D.}~\bibnamefont {Ciampini}}, \bibinfo {author}
  {\bibfnamefont {O.}~\bibnamefont {Morsch}}, \ and\ \bibinfo {author}
  {\bibfnamefont {E.}~\bibnamefont {Arimondo}},\ }\href {\doibase
  10.1103/PhysRevLett.100.040404} {\bibfield  {journal} {\bibinfo  {journal}
  {Phys. Rev. Lett.}\ }\textbf {\bibinfo {volume} {100}},\ \bibinfo {pages}
  {040404} (\bibinfo {year} {2008})}\BibitemShut {NoStop}%
\bibitem [{\citenamefont {Struck}\ \emph {et~al.}(2011)\citenamefont {Struck},
  \citenamefont {{\"O}lschl{\"a}ger}, \citenamefont {Targat}, \citenamefont
  {Soltan-Panahi}, \citenamefont {Eckardt}, \citenamefont {Lewenstein},
  \citenamefont {Windpassinger},\ and\ \citenamefont
  {Sengstock}}]{Struck:2011}%
  \BibitemOpen
  \bibfield  {author} {\bibinfo {author} {\bibfnamefont {J.}~\bibnamefont
  {Struck}}, \bibinfo {author} {\bibfnamefont {C.}~\bibnamefont
  {{\"O}lschl{\"a}ger}}, \bibinfo {author} {\bibfnamefont {R.~L.}\ \bibnamefont
  {Targat}}, \bibinfo {author} {\bibfnamefont {P.}~\bibnamefont
  {Soltan-Panahi}}, \bibinfo {author} {\bibfnamefont {A.}~\bibnamefont
  {Eckardt}}, \bibinfo {author} {\bibfnamefont {M.}~\bibnamefont {Lewenstein}},
  \bibinfo {author} {\bibfnamefont {P.}~\bibnamefont {Windpassinger}}, \ and\
  \bibinfo {author} {\bibfnamefont {K.}~\bibnamefont {Sengstock}},\ }\href
  {\doibase 10.1126/science.1207239} {\bibfield  {journal} {\bibinfo  {journal}
  {Science}\ }\textbf {\bibinfo {volume} {333}},\ \bibinfo {pages} {996}
  (\bibinfo {year} {2011})}\BibitemShut {NoStop}%
\bibitem [{\citenamefont {Struck}\ \emph {et~al.}(2012)\citenamefont {Struck},
  \citenamefont {{\"O}lschl{\"a}ger}, \citenamefont {Weinberg}, \citenamefont
  {Hauke}, \citenamefont {Simonet}, \citenamefont {Eckardt}, \citenamefont
  {Lewenstein}, \citenamefont {Sengstock},\ and\ \citenamefont
  {Windpassinger}}]{Struck:2012gc}%
  \BibitemOpen
  \bibfield  {author} {\bibinfo {author} {\bibfnamefont {J.}~\bibnamefont
  {Struck}}, \bibinfo {author} {\bibfnamefont {C.}~\bibnamefont
  {{\"O}lschl{\"a}ger}}, \bibinfo {author} {\bibfnamefont {M.}~\bibnamefont
  {Weinberg}}, \bibinfo {author} {\bibfnamefont {P.}~\bibnamefont {Hauke}},
  \bibinfo {author} {\bibfnamefont {J.}~\bibnamefont {Simonet}}, \bibinfo
  {author} {\bibfnamefont {A.}~\bibnamefont {Eckardt}}, \bibinfo {author}
  {\bibfnamefont {M.}~\bibnamefont {Lewenstein}}, \bibinfo {author}
  {\bibfnamefont {K.}~\bibnamefont {Sengstock}}, \ and\ \bibinfo {author}
  {\bibfnamefont {P.}~\bibnamefont {Windpassinger}},\ }\href {\doibase
  10.1103/PhysRevLett.108.225304} {\bibfield  {journal} {\bibinfo  {journal}
  {Phys. Rev. Lett.}\ }\textbf {\bibinfo {volume} {108}},\ \bibinfo {pages}
  {225304} (\bibinfo {year} {2012})}\BibitemShut {NoStop}%
\bibitem [{\citenamefont {Hauke}\ \emph {et~al.}(2012)\citenamefont {Hauke},
  \citenamefont {Tieleman}, \citenamefont {Celi}, \citenamefont
  {{\"O}lschl{\"a}ger}, \citenamefont {Simonet}, \citenamefont {Struck},
  \citenamefont {Weinberg}, \citenamefont {Windpassinger}, \citenamefont
  {Sengstock}, \citenamefont {Lewenstein},\ and\ \citenamefont
  {Eckardt}}]{Hauke:2012dh}%
  \BibitemOpen
  \bibfield  {author} {\bibinfo {author} {\bibfnamefont {P.}~\bibnamefont
  {Hauke}}, \bibinfo {author} {\bibfnamefont {O.}~\bibnamefont {Tieleman}},
  \bibinfo {author} {\bibfnamefont {A.}~\bibnamefont {Celi}}, \bibinfo {author}
  {\bibfnamefont {C.}~\bibnamefont {{\"O}lschl{\"a}ger}}, \bibinfo {author}
  {\bibfnamefont {J.}~\bibnamefont {Simonet}}, \bibinfo {author} {\bibfnamefont
  {J.}~\bibnamefont {Struck}}, \bibinfo {author} {\bibfnamefont
  {M.}~\bibnamefont {Weinberg}}, \bibinfo {author} {\bibfnamefont
  {P.}~\bibnamefont {Windpassinger}}, \bibinfo {author} {\bibfnamefont
  {K.}~\bibnamefont {Sengstock}}, \bibinfo {author} {\bibfnamefont
  {M.}~\bibnamefont {Lewenstein}}, \ and\ \bibinfo {author} {\bibfnamefont
  {A.}~\bibnamefont {Eckardt}},\ }\href {\doibase
  10.1103/PhysRevLett.109.145301} {\bibfield  {journal} {\bibinfo  {journal}
  {Phys. Rev. Lett.}\ }\textbf {\bibinfo {volume} {109}},\ \bibinfo {pages}
  {145301} (\bibinfo {year} {2012})}\BibitemShut {NoStop}%
\bibitem [{\citenamefont {Struck}\ \emph {et~al.}(2013)\citenamefont {Struck},
  \citenamefont {Weinberg}, \citenamefont {{\"O}lschl{\"a}ger}, \citenamefont
  {Windpassinger}, \citenamefont {Simonet}, \citenamefont {Sengstock},
  \citenamefont {H{\"o}ppner}, \citenamefont {Hauke}, \citenamefont {Eckardt},
  \citenamefont {Lewenstein},\ and\ \citenamefont {Mathey}}]{Struck:2013ar}%
  \BibitemOpen
  \bibfield  {author} {\bibinfo {author} {\bibfnamefont {J.}~\bibnamefont
  {Struck}}, \bibinfo {author} {\bibfnamefont {M.}~\bibnamefont {Weinberg}},
  \bibinfo {author} {\bibfnamefont {C.}~\bibnamefont {{\"O}lschl{\"a}ger}},
  \bibinfo {author} {\bibfnamefont {P.}~\bibnamefont {Windpassinger}}, \bibinfo
  {author} {\bibfnamefont {J.}~\bibnamefont {Simonet}}, \bibinfo {author}
  {\bibfnamefont {K.}~\bibnamefont {Sengstock}}, \bibinfo {author}
  {\bibfnamefont {R.}~\bibnamefont {H{\"o}ppner}}, \bibinfo {author}
  {\bibfnamefont {P.}~\bibnamefont {Hauke}}, \bibinfo {author} {\bibfnamefont
  {A.}~\bibnamefont {Eckardt}}, \bibinfo {author} {\bibfnamefont
  {M.}~\bibnamefont {Lewenstein}}, \ and\ \bibinfo {author} {\bibfnamefont
  {L.}~\bibnamefont {Mathey}},\ }\href {\doibase 10.1038/nphys2750} {\bibfield
  {journal} {\bibinfo  {journal} {Nat. Phys.}\ }\textbf {\bibinfo {volume}
  {9}},\ \bibinfo {pages} {738} (\bibinfo {year} {2013})}\BibitemShut {NoStop}%
\bibitem [{\citenamefont {{Jotzu}}\ \emph {et~al.}(2014)\citenamefont
  {{Jotzu}}, \citenamefont {{Messer}}, \citenamefont {{Desbuquois}},
  \citenamefont {{Lebrat}}, \citenamefont {{Uehlinger}}, \citenamefont
  {{Greif}},\ and\ \citenamefont {{Esslinger}}}]{Jotzu:2014}%
  \BibitemOpen
  \bibfield  {author} {\bibinfo {author} {\bibfnamefont {G.}~\bibnamefont
  {{Jotzu}}}, \bibinfo {author} {\bibfnamefont {M.}~\bibnamefont {{Messer}}},
  \bibinfo {author} {\bibfnamefont {R.}~\bibnamefont {{Desbuquois}}}, \bibinfo
  {author} {\bibfnamefont {M.}~\bibnamefont {{Lebrat}}}, \bibinfo {author}
  {\bibfnamefont {T.}~\bibnamefont {{Uehlinger}}}, \bibinfo {author}
  {\bibfnamefont {D.}~\bibnamefont {{Greif}}}, \ and\ \bibinfo {author}
  {\bibfnamefont {T.}~\bibnamefont {{Esslinger}}},\ }\href
  {http://arxiv.org/abs/1406.7874} {\bibfield  {journal} {\bibinfo  {journal}
  {ArXiv e-prints}\ } (\bibinfo {year} {2014})},\ \Eprint
  {http://arxiv.org/abs/1406.7874} {arXiv:1406.7874} \BibitemShut {NoStop}%
\bibitem [{\citenamefont {{Zhang}}\ and\ \citenamefont
  {{Zhou}}(2014)}]{Zhang:2014}%
  \BibitemOpen
  \bibfield  {author} {\bibinfo {author} {\bibfnamefont {S.-L.}\ \bibnamefont
  {{Zhang}}}\ and\ \bibinfo {author} {\bibfnamefont {Q.}~\bibnamefont
  {{Zhou}}},\ }\href {http://arxiv.org/abs/1403.0210} {\bibfield  {journal}
  {\bibinfo  {journal} {ArXiv e-prints}\ } (\bibinfo {year} {2014})},\ \Eprint
  {http://arxiv.org/abs/1403.0210} {arXiv:1403.0210} \BibitemShut {NoStop}%
\bibitem [{\citenamefont {{Goldman}}\ and\ \citenamefont
  {{Dalibard}}(2014)}]{Goldman:2014}%
  \BibitemOpen
  \bibfield  {author} {\bibinfo {author} {\bibfnamefont {N.}~\bibnamefont
  {{Goldman}}}\ and\ \bibinfo {author} {\bibfnamefont {J.}~\bibnamefont
  {{Dalibard}}},\ }\href {http://arxiv.org/abs/1404.4373} {\bibfield  {journal}
  {\bibinfo  {journal} {ArXiv e-prints}\ } (\bibinfo {year} {2014})},\ \Eprint
  {http://arxiv.org/abs/1404.4373} {arXiv:1404.4373} \BibitemShut {NoStop}%
\bibitem [{\citenamefont {Baur}\ \emph {et~al.}(2014)\citenamefont {Baur},
  \citenamefont {Schleier-Smith},\ and\ \citenamefont {Cooper}}]{Baur:2014}%
  \BibitemOpen
  \bibfield  {author} {\bibinfo {author} {\bibfnamefont {S.~K.}\ \bibnamefont
  {Baur}}, \bibinfo {author} {\bibfnamefont {M.~H.}\ \bibnamefont
  {Schleier-Smith}}, \ and\ \bibinfo {author} {\bibfnamefont {N.~R.}\
  \bibnamefont {Cooper}},\ }\href {\doibase 10.1103/PhysRevA.89.051605}
  {\bibfield  {journal} {\bibinfo  {journal} {Phys. Rev. A}\ }\textbf {\bibinfo
  {volume} {89}},\ \bibinfo {pages} {051605} (\bibinfo {year}
  {2014})}\BibitemShut {NoStop}%
\bibitem [{\citenamefont {Kosior}\ and\ \citenamefont
  {Sacha}(2014)}]{Kosior:2014}%
  \BibitemOpen
  \bibfield  {author} {\bibinfo {author} {\bibfnamefont {A.}~\bibnamefont
  {Kosior}}\ and\ \bibinfo {author} {\bibfnamefont {K.}~\bibnamefont {Sacha}},\
  }\href {\doibase 10.1209/0295-5075/107/26006} {\bibfield  {journal} {\bibinfo
   {journal} {Europhys. Lett.}\ }\textbf {\bibinfo {volume} {107}},\ \bibinfo
  {pages} {26006} (\bibinfo {year} {2014})}\BibitemShut {NoStop}%
\bibitem [{\citenamefont {Zhang}\ \emph {et~al.}(2013)\citenamefont {Zhang},
  \citenamefont {Chen},\ and\ \citenamefont {Zhang}}]{Zhang:2013es}%
  \BibitemOpen
  \bibfield  {author} {\bibinfo {author} {\bibfnamefont {Y.}~\bibnamefont
  {Zhang}}, \bibinfo {author} {\bibfnamefont {G.}~\bibnamefont {Chen}}, \ and\
  \bibinfo {author} {\bibfnamefont {C.}~\bibnamefont {Zhang}},\ }\href
  {\doibase 10.1038/srep01937} {\bibfield  {journal} {\bibinfo  {journal} {Sci.
  Rep.}\ }\textbf {\bibinfo {volume} {3}},\ \bibinfo {pages} {1937} (\bibinfo
  {year} {2013})}\BibitemShut {NoStop}%
\bibitem [{\citenamefont {Shirley}(1965)}]{Shirley:1965cy}%
  \BibitemOpen
  \bibfield  {author} {\bibinfo {author} {\bibfnamefont {J.~H.}\ \bibnamefont
  {Shirley}},\ }\href {\doibase 10.1103/PhysRev.138.B979} {\bibfield  {journal}
  {\bibinfo  {journal} {Phys. Rev.}\ }\textbf {\bibinfo {volume} {138}},\
  \bibinfo {pages} {979} (\bibinfo {year} {1965})}\BibitemShut {NoStop}%
\bibitem [{\citenamefont {Sambe}(1973)}]{Sambe:1973hi}%
  \BibitemOpen
  \bibfield  {author} {\bibinfo {author} {\bibfnamefont {H.}~\bibnamefont
  {Sambe}},\ }\href {\doibase 10.1103/PhysRevA.7.2203} {\bibfield  {journal}
  {\bibinfo  {journal} {Phys. Rev. A}\ }\textbf {\bibinfo {volume} {7}},\
  \bibinfo {pages} {2203} (\bibinfo {year} {1973})}\BibitemShut {NoStop}%
\bibitem [{\citenamefont {Hemmerich}(2010)}]{Hemmerich:2010fh}%
  \BibitemOpen
  \bibfield  {author} {\bibinfo {author} {\bibfnamefont {A.}~\bibnamefont
  {Hemmerich}},\ }\href {\doibase 10.1103/PhysRevA.81.063626} {\bibfield
  {journal} {\bibinfo  {journal} {Phys. Rev. A}\ }\textbf {\bibinfo {volume}
  {81}},\ \bibinfo {pages} {063626} (\bibinfo {year} {2010})}\BibitemShut
  {NoStop}%
\bibitem [{\citenamefont {Arlinghaus}\ and\ \citenamefont
  {Holthaus}(2010)}]{Arlinghaus:2010bz}%
  \BibitemOpen
  \bibfield  {author} {\bibinfo {author} {\bibfnamefont {S.}~\bibnamefont
  {Arlinghaus}}\ and\ \bibinfo {author} {\bibfnamefont {M.}~\bibnamefont
  {Holthaus}},\ }\href {\doibase 10.1103/PhysRevA.81.063612} {\bibfield
  {journal} {\bibinfo  {journal} {Phys. Rev. A}\ }\textbf {\bibinfo {volume}
  {81}},\ \bibinfo {pages} {063612} (\bibinfo {year} {2010})}\BibitemShut
  {NoStop}%
\bibitem [{\citenamefont {Poletti}\ and\ \citenamefont
  {Kollath}(2011)}]{Poletti:2011}%
  \BibitemOpen
  \bibfield  {author} {\bibinfo {author} {\bibfnamefont {D.}~\bibnamefont
  {Poletti}}\ and\ \bibinfo {author} {\bibfnamefont {C.}~\bibnamefont
  {Kollath}},\ }\href {\doibase 10.1103/PhysRevA.84.013615} {\bibfield
  {journal} {\bibinfo  {journal} {Phys. Rev. A}\ }\textbf {\bibinfo {volume}
  {84}},\ \bibinfo {pages} {013615} (\bibinfo {year} {2011})}\BibitemShut
  {NoStop}%
\bibitem [{\citenamefont {Cole}\ \emph {et~al.}(2012)\citenamefont {Cole},
  \citenamefont {Zhang}, \citenamefont {Paramekanti},\ and\ \citenamefont
  {Trivedi}}]{Cole:2012}%
  \BibitemOpen
  \bibfield  {author} {\bibinfo {author} {\bibfnamefont {W.~S.}\ \bibnamefont
  {Cole}}, \bibinfo {author} {\bibfnamefont {S.}~\bibnamefont {Zhang}},
  \bibinfo {author} {\bibfnamefont {A.}~\bibnamefont {Paramekanti}}, \ and\
  \bibinfo {author} {\bibfnamefont {N.}~\bibnamefont {Trivedi}},\ }\href
  {\doibase 10.1103/PhysRevLett.109.085302} {\bibfield  {journal} {\bibinfo
  {journal} {Phys. Rev. Lett.}\ }\textbf {\bibinfo {volume} {109}},\ \bibinfo
  {pages} {085302} (\bibinfo {year} {2012})}\BibitemShut {NoStop}%
\bibitem [{\citenamefont {Cai}\ \emph {et~al.}(2012)\citenamefont {Cai},
  \citenamefont {Zhou},\ and\ \citenamefont {Wu}}]{Zi:2012}%
  \BibitemOpen
  \bibfield  {author} {\bibinfo {author} {\bibfnamefont {Z.}~\bibnamefont
  {Cai}}, \bibinfo {author} {\bibfnamefont {X.}~\bibnamefont {Zhou}}, \ and\
  \bibinfo {author} {\bibfnamefont {C.}~\bibnamefont {Wu}},\ }\href {\doibase
  10.1103/PhysRevA.85.061605} {\bibfield  {journal} {\bibinfo  {journal} {Phys.
  Rev. A}\ }\textbf {\bibinfo {volume} {85}},\ \bibinfo {pages} {061605}
  (\bibinfo {year} {2012})}\BibitemShut {NoStop}%
\bibitem [{\citenamefont {Radi\ifmmode~\acute{c}\else \'{c}\fi{}}\ \emph
  {et~al.}(2012)\citenamefont {Radi\ifmmode~\acute{c}\else \'{c}\fi{}},
  \citenamefont {Di~Ciolo}, \citenamefont {Sun},\ and\ \citenamefont
  {Galitski}}]{Radic:2012}%
  \BibitemOpen
  \bibfield  {author} {\bibinfo {author} {\bibfnamefont {J.}~\bibnamefont
  {Radi\ifmmode~\acute{c}\else \'{c}\fi{}}}, \bibinfo {author} {\bibfnamefont
  {A.}~\bibnamefont {Di~Ciolo}}, \bibinfo {author} {\bibfnamefont
  {K.}~\bibnamefont {Sun}}, \ and\ \bibinfo {author} {\bibfnamefont
  {V.}~\bibnamefont {Galitski}},\ }\href {\doibase
  10.1103/PhysRevLett.109.085303} {\bibfield  {journal} {\bibinfo  {journal}
  {Phys. Rev. Lett.}\ }\textbf {\bibinfo {volume} {109}},\ \bibinfo {pages}
  {085303} (\bibinfo {year} {2012})}\BibitemShut {NoStop}%
\bibitem [{\citenamefont {Xu}\ \emph {et~al.}(2014)\citenamefont {Xu},
  \citenamefont {Cole},\ and\ \citenamefont {Zhang}}]{Xu:2014}%
  \BibitemOpen
  \bibfield  {author} {\bibinfo {author} {\bibfnamefont {Z.}~\bibnamefont
  {Xu}}, \bibinfo {author} {\bibfnamefont {W.~S.}\ \bibnamefont {Cole}}, \ and\
  \bibinfo {author} {\bibfnamefont {S.}~\bibnamefont {Zhang}},\ }\href
  {\doibase 10.1103/PhysRevA.89.051604} {\bibfield  {journal} {\bibinfo
  {journal} {Phys. Rev. A}\ }\textbf {\bibinfo {volume} {89}},\ \bibinfo
  {pages} {051604} (\bibinfo {year} {2014})}\BibitemShut {NoStop}%
\end{thebibliography}
\end{document}